\documentclass [12pt]{article} 
\usepackage{latexsym}
\newcommand{\KK}{\mathsf{K}}
\newcommand{\LL}{\mathcal{L}}
\newcommand{\HH}{\mathcal{H}} 
\newcommand{\TT}{\mathrm{Tr}} 
\newcommand{\TP}{\mbox{\boldmath $\sqcap$}} 
\newcommand{\GG}{\mathcal{G}} 

\title{Constraints in  Quantum Geometrodynamics} 

\author{
Adrian P. Gentle\thanks{Department of Mathematics, University of Southern Indiana, 
Evansville, IN 47712 and 
Theoretical Division, Los Alamos National Laboratory, Los Alamos, NM 87545},
Nathan D. George\thanks{DAMTP, University of Cambridge, Cambridge CB3 0WA, United 
Kingdom and
Theoretical Division, Los Alamos National Laboratory, Los Alamos, NM 87545},
Arkady Kheyfets\thanks{Department of Mathematics, North Carolina State University, 
Raleigh, NC 27695} and \\
Warner A. Miller\thanks{Department of Physics, Florida Atlantic University, Boca Raton,
FL 33431 and 
Theoretical Division, Los Alamos National Laboratory, Los Alamos, NM 87545} 
}

\date{\today}

\begin{document} 
 
\maketitle

\begin{abstract} 
  
  We compare different treatments of the constraints in canonical
  quantum gravity. The standard approach on the superspace of
  3--geometries treats the constraints as the sole carriers of the
  dynamic content of the theory, thus rendering the traditional
  dynamical equations obsolete.  Quantization of the constraints in
  both the Dirac and ADM square root Hamiltonian approaches leads to
  the well known problems of time evolution.  These problems of time
  are of both an interpretational and technical nature. In contrast,
  the geometrodynamic quantization procedure on the superspace of the
  true dynamical variables separates the issues of quantization from
  the enforcement of the constraints. The resulting theory takes into
  account states that are off-shell with respect to the constraints,
  and thus avoids the problems of time.  We develop, for the first
  time, the geometrodynamic quantization formalism in a general
  setting and show that it retains all essential features previously
  illustrated in the context of homogeneous cosmologies.

\end{abstract}


\section{Introduction.} 
\label{I}

The standard approach to the canonical quantization of
gravity\cite{Isham99,Kuc93} is based on the classical dynamic
picture of the evolving 3--geometry of a slicing of a spacetime
manifold.  The slicing is essentially a reference foliation of the
spacetime manifold (endowed with a 4--geometry) with respect to which
canonical variables are assigned.  It is usually parametrized by a
time coordinate $t$ and tied to the enveloping spacetime by the lapse
function $N$ and the shift vector $N^i$.  The canonical variables are
the 3--metric $g_{ik}$ on a spatial slices of the
foliation induced by the spacetime 4--metric, and the matrix
$\pi^{ik}$ of their canonically conjugate momenta. The latter is
related to the extrinsic curvature of $\Sigma$ when it is considered
to be embedded in the spacetime.

The customary variational procedure, applied to the Hilbert action
expressed in terms of the canonical variables, produces Hamilton
equations which describe the time evolution of the canonical
variables.  The Hamiltonian is given by $N {\cal H} + N_i {\cal H}^i$,
where ${\cal H}$ and ${\cal H}^i$ are functions of the canonical
variables and their spatial derivatives.  The procedure is not
extended to the derivation of the Hamilton--Jacobi equation in the
usual manner, as such an equation with the chosen
set of canonical variables\cite{Mis57} appears to become meaningless
when the general covariance of the theory is implemented 
(cf.~ section \ref{V} for a more precise description).

Alternatively, general covariance can be introduced in the variational
principle from the outset, by requiring that the action is invariant
with respect to variations of the lapse and shift.  This leads to the
constraint equations (to simplify notations, we omit indices on
components of $g$ and $\pi$ in all equations in this section)
\begin{equation} 
{\cal H}(g, \pi ; x) = 0
\end{equation}   
and 
\begin{equation} 
{\cal H}^i(g, \pi ; x) = 0
\end{equation}   
which are imposed on the canonical variables of each slice.  An
important feature of general relativity is that its dynamics is fully
constrained.  It can be shown that if the geometry of spacetime is
such that the constraints are satisfied on all slices of all spatial
foliations of spacetime, then the canonical variables necessarily
satisfy the Hamilton evolution equations.  This feature is often
referred to as a key property of general relativity\cite{Isham99} and
is used to argue that the entire theory is encoded in the
constraints, with the conclusion that the Hamilton equations are
redundant and can be ignored in dynamical considerations.
Substitution of $\delta S/\delta g$ in the place of $p$ in the
constraint equations leads to a new set of equations
\begin{equation} 
{\cal H}\left( g, {\delta S\over\delta g}; x\right) = 0
\end{equation}   
and 
\begin{equation} 
{\cal H}^i\left( g, {\delta S\over\delta g}; x\right) = 0,
\end{equation}  
the first of which is considered to be the Hamilton--Jacobi equation
(see section \ref{V}).  Arguments appealing to the variational
principle on the superspace of 3--geometries support this assertion.
Detailed arguments and the interpretation of other equations may be
found elsewhere\cite{MTW70}.

Dirac's procedure of canonical gravity quantization is based directly
on this Hamilton--Jacobi equation, and produces a quantum theory that
consists of the Wheeler--DeWitt equation together with commutation 
relations imposed on all canonical variables.

The ADM square root quantization procedure is also based entirely on
constraints, but in this procedure the set of canonical variables is
split into two subsets: the embedding variables (four of them
altogether; one slicing parameter $\Omega$ and three coordinatization
parameters $\alpha$) and the true dynamical variables $\beta$ (two of
them)\cite{Kuc92,KheMil94a,KheMil96}.  The constraints are then solved
with respect to the momenta conjugate to the embedding variables.
After substituting $\delta S/\delta\Omega$, $\delta S/\delta\alpha$
for $p_\Omega$, $p_\alpha$, (where $S$ is the principal Hamilton
function) one of the resulting equations (the equation for the
momentum conjugate to the slicing parameter) is identified with the
Hamilton--Jacobi equation, and its right hand side yields an
expression for a new (square root) Hamiltonian.  The quantization is
based on this equation, yielding a quantum theory that consists of the
Schr\"odinger equation and commutation relations imposed on the true
dynamical variables and their conjugate momenta.

Describing the time evolution of quantized gravitational fields
is extremely troublesome in both formulations.  Any attempt to
introduce time that can be used in a way similar to time in quantum
mechanics, or in quantum field theory on a flat background, invariably
leads to the notorious problems of time\cite{Kuc92}.  Attempts to
introduce time in a universal way externally (such as the readings of
specially designed clock) have been unsuccessful.  There are reasons
to believe that this is impossible\cite{Isham99}, whether the clock is
believed to be gravitationally defined (i.e. the readings depend only
on the variables describing the gravitational field), or a matter
clock (the readings depend on both gravitational and matter
variables).

The conceptual difficulties (such as the problem of functional
evolution and the multiple choice problem, in Kucha\v r's terminology)
emerge due to the dual nature of time parametrization in general
relativity.  If spacetime is considered as a manifold, it can be
coordinatized and sliced in an arbitrary manner.  However, this is not
sufficient for the description of geometrodynamic evolution, since
both slicing and coordinatization need to be tied to the metric of
spacetime.  The standard approach in classical geometrodynamics
involves lapse and shift. These quantites determine the slicing of
spacetime (i.e. the slicing condition) and its coordinatization.  On a
physical basis the lapse (the slicing) represents the reading of the
clock of the observer whose worldline is perpindicular to the
spacelike hypersurface. While the shift represents the displacement of
the spatial coordinates in time away from such an observer. Here, the
dynamics is determined solely interms of the constraints only if there
exists a unique spacetime metric.  This metric might not be known
until the geometrodynamic problem is solved.  While its existence is
not a problem in classical general relativity, there is, in general,
no possibility of assigning such a unique metric of spacetime in
canonical quantum gravity.

Given these problems, continuing to use the constraint equations as
the foundation of geometrodynamics becomes problematic, to say the
least.  Quantization of the dynamical picture based on the constraints
is essentially equivalent to restricting the states of the resulting
quantum systems to a ``shell'', which is determined by constraints
that are classical in origin.  An attempt to undertake a similar
action in quantum mechanics or quantum field theory would be quite
disastrous under all but very carefully selected conditions.

One way to resolve this dilemma would be to weaken the requirement of
covariance, essentially discarding it in dynamical considerations and
recovering it by imposing symmetries on solutions only to the extent
and in the sense that is allowed by dynamics.  General covariance in
the traditional sense should be recovered in the classical limit.
This requirement should determine, at least partially, what
constitutes the classical limit of quantum geometrodynamics.
 
This can be achieved if York's analysis of gravitational degrees of
freedom\cite{Yor72} is taken into account and actively implemented.
The Hamilton--Jacobi equation takes its traditional form, familiar
from classical mechanics or electrodynamics. The resulting
description remains equivalent to the one commonly accepted in
classical geometrodynamics.  However, quantization based on this new
Hamilton--Jacobi equation provides an appropriate interpretation of
the conceptual problems of time, making them quite natural statements
concerning the properties of gravity quantization.  It also seems to
avoid the technical problems of time, such as the Hilbert space and
spectral analysis problems, as it produces a Schr\"odinger equation
for the state evolution, and the Hamiltonian does not include the
square root operation.  The procedure has been described elsewhere in
the context of homogeneous quantum
cosmologies\cite{KheMil94a,KheMil96}.

In this setting time can be introduced as a slicing parametrization on
a spacetime manifold, and tied to the metric structure, without
contradiction.  The metric interpretation of time is coupled with
geometrodynamic evolution, while the true meaning of time becomes
completely determined only after the geometrodynamic evolution problem
has been solved.  In a sense, the quantum geometrodynamic
configuration and time emerge together, and the meaning of the clock
readings is influenced by the quantum gravitational
system\cite{KheMil96}.

As noted above, the geometrodynamic approach differs from the standard
one in its treatment of the constraints.  This difference is rather
subtle on the classical level and does not result in different
predictions.  On the quantum level, however, it becomes fundamental.

In previous work we have considered different aspects of
geometrodynamic quantization in particular cases of homogeneous
quantum cosmologies.  Here, we develop, for the first time, the
geometrodynamic quantization formalism in a general setting and show
that it retains all of the essential features previously illustrated
in the context of homogeneous cosmologies.

To achieve transparency of the discussion and to provide an
appropriate platform for future applications, we start by reviewing
the issue in classical geometrodynamics.  We begin from the Lagrangian
formulation, and then describe the transition to the Hamilton and 
Hamilton--Jacobi equations.

\section{Constraints on the Configuration Space of 3--Metrics.} 
\label{II} 

We start from the 3+1 Lagrangian expression for the action, obtained
from the standard Hilbert action by expressing the 4--metric in terms
of lapse $N$, shift $N_i$, and 3--metric $g_{ij}$. After eliminating
total time derivatives and total divergences\cite{MTW70}, the action
becomes
\begin{equation}
I_c = \frac{1}{16\pi} \int \left[ R - (\TT\KK )^2 + \TT (\KK^2)\right] 
N \sqrt{g}\, d^3x\, dt = \int \LL\, d^3x\, dt 
\end{equation} 
where 
\begin{equation} 
\label{lagr}
\LL = \frac{1}{16\pi}\, \left[ R - (\TT\KK )^2 + \TT (\KK^2)\right] 
N \sqrt{g}.
\end{equation} 
Following standard convention\cite{MTW70} we drop the factor of
$1/16\pi$ from the gravitational action in the remainder of this
paper.  In vacuum the equations are equivalent, and the factor can be
trivially included when matter is present.
 
In the equations above $\KK$ is used as a shorthand notation for the
extrinsic curvature $K_{ij}$, defined as
\begin{equation} 
\label{excr}
K_{ij} = \frac{1}{2N} \left[ N_{i| j} +N_{j|i} - \dot{g}_{ij}\right],
\end{equation} 
$R$ is the 3--curvature associated with the 3--metric $g_{ij}$, $g$ is
the determinant of this 3--metric, and $\dot{g}_{ij} = \frac{\partial
  g_{ij}}{\partial t}$.

Components of the 3--metric $g_{ij}$ are treated as dynamical
variables or (functional) coordinates of the geometrodynamic
configuration space.  Following the standard prescription for
transitioning from the Lagrangian description of dynamics to the
Hamiltonian one, we introduce momenta $\pi^{ij}$ conjugate to the
dynamical variables $g_{ij}$, with
\begin{equation} 
\label{M1} 
\pi^{ij} = \frac{\partial\LL}{\partial\dot{g}_{ij}} .
\end{equation} 
Computing the right hand side of this expression involves determining the 
derivatives of $K = \TT (\KK ) = g^{ij} K_{ij}$ and of $\TT (\KK^2) =
g^{im} g^{jk} K_{mj} K_{ki}$ with respect to the $\dot{g}_{nl}$ using
\begin{equation} 
\frac{\partial K_{ij}}{\partial\dot{g}_{km}} = 
-\frac{1}{2N}\, {\delta^k}_i\, {\delta^m}_j .
\end{equation} 
This computation yields 
\begin{equation} 
  \frac{\partial\TT\KK}{\partial\dot{g}_{nl}} = -\frac{1}{2N}\, 
  g^{ij}\, {\delta^n}_i\, {\delta^l}_j =  -\frac{1}{2N}\, g^{nl} 
\end{equation}
\begin{eqnarray} 
  \frac{\partial (\TT\KK^2)}{\partial\dot{g}_{nl}} &=&
-\frac{1}{2N}\, \left[ g^{im} g^{jk} {\delta^n}_m\, {\delta^l}_j\, K_{ki} + 
g^{im} g^{jk} K_{mj}\, {\delta^n}_k\, {\delta^l}_j\right]  \nonumber \\
 &=& -\frac{1}{2N}\, \left[ g^{in} g^{lk} K_{ki} + 
g^{jn} g^{lm} K_{mj}\right] = -\frac{1}{N}\, K^{nl} 
\end{eqnarray} 
which results in the expression for the momenta $\pi^{ij}$ defined by 
(\ref{M1}),
\begin{equation} 
\label{pik} 
\pi^{ij} = \sqrt{g}\, \left( K g^{ij} - K^{ij}\right).
\end{equation} 
In what follows we use notations $\Pi = \pi^{ij}$ and $\TP = \TT\Pi$. The 
last equation implies 
\begin{eqnarray} 
  \TP & = & \sqrt{g}\, (3K - K) = 2 \sqrt{g} K \\
  \label{ktp}
  K^{ij} & = & \frac{1}{\sqrt{g}}\, \left(\frac{1}{2}\TP g^{ij} - \pi^{ij}\right) \\
  \TT\KK^2 & = & \frac{1}{g}\, \left(\TT\Pi^2 - \frac{\TP^2}{4}\right) 
\end{eqnarray} 
which allows one to express $\LL$ as a function of $g_{ij}$ and $\pi^{ij}$ 
only, giving
\begin{equation} 
\LL = N\, \left[ g^{\frac{1}{2}} R - g^{-\frac{1}{2}} 
\left(\frac{\TP^2}{2} - \TT\Pi^2\right)\right] .
\end{equation} 
The standard definition of the Hamiltonian 
\begin{equation} 
\label{hwdw}
\HH_{WDW} = \pi^{ij} \dot{g}_{ij} - \LL\qquad 
\hbox{\rm (Mod total divergence)}  
\end{equation} 
can be transformed to express $\HH_{WDW}$ in terms of $g_{ij}$ and the
conjugate momenta.  In order to achieve this, we use
\begin{equation} 
\dot{g}_{ij} = N_{i|j} + N_{j|i} - 2 N K_{ij} 
\end{equation} 
which allows us to write the first term of (\ref{hwdw}) as 
\begin{equation} 
\begin{array}{rl} 
\pi^{ij} \dot{g}_{ij} &= (N_{i|j} + N_{j|i}) \pi^{ij} - 
2 N \pi^{ij}K_{ij} = 2 N_{i|j} \pi^{ij} - 2 N \pi^{ij}K_{ij} \\ 
 & \\ 
 &= 2 \left( N_i \pi^{ij}\right)_{|j} - 
2 N_i \left( \pi^{ij}\right)_{|j} 
- 2 N \pi^{ij}K_{ij} .
\end{array}  
\end{equation} 
The first term in this expression is the total covariant divergence of
a vector density.  It can be expressed as the total divergence of a
vector field,
\begin{equation}
\left( N_i \pi^{ij}\right)_{|j} = \sqrt{g}\, 
\left[ N_i (K g^{ij} - K^{ij})\right]_{|j} = 
\left[ \sqrt{g}\, N_i (K g^{ij} - K^{ij})\right]_{,j} 
\end{equation} 
and removed from the final expression for the Hamiltonian. The third
term, after substituting $K_{ij}$ as given by (\ref{ktp}), reduces to
\begin{equation} 
2 N \pi^{ij}K_{ij} = 2 g^{-\frac{1}{2}} 
N \left(\frac{\TP^2}{2} - \TT\Pi^2 \right),
\end{equation} 
and the expression for $\HH_{WDW}$ takes the form
\begin{equation} 
\HH_{WDW} = N_i \left( - 2 \pi^{ij}\right)_{|j} + N \left[ 
g^{-\frac{1}{2}} \left(\TT\Pi^2 - \frac{\TP^2}{2} \right) - 
g^{\frac{1}{2}} R\right] .
\end{equation} 
It is common practice to write the Hamiltonian $\HH_{WDW}$ in the form
\begin{equation} 
\label{hamg}
\HH_{WDW} = N\, \HH + N_i\, \HH^i 
\end{equation} 
where 
\begin{equation}
\label{suphg} 
\HH = g^{-\frac{1}{2}} \left(\TT\Pi^2 - \frac{\TP^2}{2} \right) - 
g^{\frac{1}{2}} R 
\end{equation} 
is called the superhamiltonian, and 
\begin{equation} 
\label{supmg}
\HH^i = -2\, \left(\pi^{ij}\right)_{|j} 
\end{equation} 
are the supermomenta.  The action in Hamiltonian form can thus be
written as
\begin{eqnarray} 
\label{icosm}
I_c &=& \int \left(\pi^{ij} \dot{g}_{ij} - \HH_{WDW}\right)\, d^3x\,
dt\nonumber \\ 
&=& \int \left(\pi^{ij} \dot{g}_{ij} - N\, \HH - N_i\, \HH^i\right)\, 
d^3x\, dt .
\end{eqnarray} 
Variations of this action with respect to $\pi^{ij}$ and $g_{ij}$ 
produces the Hamilton equations 
\begin{eqnarray} 
\label{ham1}
\dot{g}_{ij} &=& \frac{\partial\HH_{WDW}}{\partial\pi^{ij}} \\
\dot{\pi}^{ij} &=& -\frac{\partial\HH_{WDW}}{\partial g_{ij}},
\end{eqnarray} 
while variations of the shift and lapse yield the superhamiltonian and
supermomenta constraint equations
\begin{equation} 
\label{hcns}
\HH (g_{ij}, \pi^{ij}) = 0 
\end{equation} 
\begin{equation} 
\label{hicns}
\HH^i(g_{ij}, \pi^{ij}) = 0,
\end{equation}
which will be discussed in more detail after we develop convenient
notations and formal machinery to handle the relevant questions.  It
is important that these equations (\ref{ham1}) can be obtained by
inverting the kinematic relations (\ref{pik}); therefore, they are 
independent of the Hamiltonian dynamics.

\section{Transformation of Variables.} 
\label{III} 

It is often convenient to replace $g_{ij}$ with another set of
variables (functions) which form the configuration space.  For
instance, in general analysis of the initial data problem or
gravitational degrees of freedom the variables are split into the true
dynamical variables, the scale factor and the gravitomagnetic vector
(three components) that cannot be identified with components of the
3--metric $g_{ij}$. A similar parametrization is used in setting up
the problems of homogeneous cosmologies\cite{RS75}. In general, the
$g_{ij}$ are expressed as
\begin{equation} 
\label{vtr}  
g_{ij} = g_{ij}(q_A) 
\end{equation} 
or 
\begin{equation} 
g_{ij} = g_{ij}(q_A, x^i) 
\end{equation} 
where $q_A = q_A(x^i, t)$ are assumed to be independent, and $A= 1,
\ldots, n_q$ with $n_q \le 6$. In the generic case $n_q = 6$ and all
components of shift are present in the spacetime metric.  If this is
not the case, some symmetries have been used to fix the form of the
metric, which typically leads to the loss of covariance.  Some of the
supermomenta constraints might be lost or not be independent. It is
hard to list and consider all possibilities, but as a rule these
degenerate cases do not cause essential troubles in any particular
case.  In what follows we will be interested mostly in the generic,
non-degenerate case, although most of the conclusions will also be
true for degenerate cases.

The transition from the Lagrangian to the Hamiltonian action begins
with the Lagrangian (\ref{lagr}).  The extrinsic curvature is given by
(\ref{excr}), except now $\dot{g}_{ij}$ is given by
\begin{equation} 
\label{jac} 
\dot{g}_{ij} = \frac{\partial g_{ij}}{\partial q_A}\, \dot{q}_A = 
{M^A}_{ij} \dot{q}_A 
\end{equation} 
which results in the expression for $K_{ij}$ 
\begin{equation} 
\label{kkq}
K_{ij} = \frac{1}{2N}\, \left[ N_{i|j} + N_{j|i} - {M^A}_{ij} \dot{q}_A 
\right] 
\end{equation} 
and 
\begin{equation} 
\label{kqd} 
\frac{\partial K_{ij}}{\partial\dot{q}_A} = -\frac{1}{2N}\, {M^A}_{ij},
\end{equation} 
where ${M^A}_{ij}$ is a {$3\times 3$} symmetric matrix for each
value of $A$.
Alternatively, one can narrow it down to six components of $g_{ij}$
such that $i \leq j$ (right upper triangle) and consider pairs $(ij)$
as collective indices, in which case this matrix becomes the Jacobian
of transformation between $q_A$ and $g_{ij}, i\leq j$. The assumption
that the variables $q_A$ are independent implies that the rank of this
matrix is equal to the number of $q_A$ (less or equal to six). In the
generic cases this rank is equal to six, which means that the system
of equations (\ref{jac}) can be solved, providing expressions for
$\dot{q}_A$ in terms of $\dot{g}_{ij}$.
   
Equations (\ref{kqd}) allow us to calculate the derivatives of 
\begin{eqnarray} 
  \TT\KK &=& g^{ij} K_{ij}  \\
  \TT (\KK^2) &=& g^{im}\, g^{jk}\, K_{mj}\, K_{ki} 
\end{eqnarray} 
with respect to $\dot{q}_A$, which yields 
\begin{eqnarray} 
\frac{\partial\TT\KK}{\partial\dot{q}_A} & = & 
-\frac{1}{2N}\, g^{ij} {M^A}_{ij} \\
\frac{\partial\left(\TT (\KK^2)\right)}{\partial\dot{q}_A}
& = & 
g^{im}\, g^{ik}\, \left( -\frac{1}{2N}\, {M^A}_{mj}\right)\, K_{ki}
 + g^{im}\, g^{ik}\, K_{mj}\, \left( -\frac{1}{2N}\,
  {M^A}_{ki}\right) \nonumber  \\ 
& = & -\frac{1}{2N}\, \left[ K^{jm}\, {M^A}_{mj} + K^{ik}\, 
  {M^A}_{ki}\right] \nonumber  \\
& = & -\frac{1}{N}\, K^{ij}\, {M^A}_{ij} .
\end{eqnarray} 
The momentum $p^A$ conjugate to the variable $q_A$ is given by 
\begin{eqnarray} 
\label{momv}
\label{pvr}
p^A\  =\  \frac{\partial\LL}{\partial\dot{q}_A} & = &
\sqrt{g} \left[ K\, g^{ij} {M^A}_{ij} - K^{ij} {M^A}_{ij}\right]
\nonumber \\
 & = & \sqrt{g} \left[ K\, g^{ij} - K^{ij}\right]\, {M^A}_{ij} 
\end{eqnarray}
and comparison of this expression with (\ref{pik}) yields the very
useful relations
\begin{equation} 
\label{tran}
p^A = \pi^{ij} {M^A}_{ij} .
\end{equation} 
The Lagrangian $\LL$ (\ref{lagr}), with $\KK$ given by (\ref{kkq}),
becomes a function of the new variables
\begin{equation} 
\LL = \LL (q_A, \dot{q}_A, N, N_i) , 
\end{equation} 
and the transition to the Hamiltonian action can proceed in the
standard way, expressing the Hamiltonian
\begin{equation} 
\widetilde\HH = p^A \dot{q}_A - \LL ,
\end{equation}   
in terms of canonical variables $q_A$, $p^A$.  This procedure
presumably results in an expression for the Hamiltonian of the same
type as above:
\begin{equation} 
\label{hamex}
\widetilde\HH = N\, \HH (q_A, p^A) + N_i\, \HH^i(q_A, p^A) .
\end{equation} 
This is obvious in the generic case, when the rank of ${M^A}_{ij}$ is
maximal (six), since (\ref{hamex}) can then be obtained by inverting
(\ref{vtr}) and (\ref{tran}) and substituting the expressions for
$\pi^{ij} = \pi^{ij}(q_A, p^A)$, together with (\ref{vtr}), into
(\ref{hamg})--(\ref{supmg}). The resulting Hamiltonian $\widetilde\HH$
depends on the new variables $q_A, p^A$, and inverting (\ref{vtr}),
followed by the substitution of $q_A = q_A(g_{ij})$ and (\ref{tran})
into this new Hamiltonian, yields
\begin{equation} 
\widetilde\HH\left( q_A(g_{ij}), p^A(g_{ij}, \pi^{ij}), N, N_i\right) = 
\HH_{WDW}\left( g_{ij}, \pi^{ij}, N, N_i\right),
\end{equation} 
which implies 
\begin{equation} 
\dot{g}_{ij} = {M^A}_{ij}\, \dot{q}_A = 
\frac{\partial\HH_{WDW}}{\partial\pi^{ij}} = 
\frac{\partial p^A}{\partial\pi^{ij}}\, 
\frac{\partial\widetilde\HH}{\partial p^A} = 
{M^A}_{ij}\, \frac{\partial\widetilde\HH}{\partial p^A}.
\end{equation} 
Thus
\begin{equation} 
{M^A}_{ij}\, \left(\dot{q}_A - 
\frac{\partial\widetilde\HH}{\partial p^A}\right) = 0,
\end{equation}  
and this system of equations has the unique solution 
\begin{equation} 
\label{qav}
\dot{q}_A = \frac{\partial\widetilde\HH}{\partial p^A} .
\end{equation} 
Just as in the previous section, these expressions for $\dot{q}_A$ are 
equivalent to the definitions of momenta $p^A$ given by (\ref{momv}). 

This logic fails when the number $n_q$ of variables $q_A$ is less than
six.  Nevertheless, the basic structure remains the same. In
particular, the Hamiltonian $\widetilde\HH\left( q_A, p^A, N,
  N_i\right)$ can be expressed by (\ref{hamex}) as a combination of
the superhamiltonian and supermomenta (neither of which depend on
shift and lapse).  Note, however, that this expression cannot be
obtained by the simple inversions described above.  Just as in the 
non-degenerate case, the Hamilton equations (\ref{qav}) remain 
equivalent to the definitions of momenta (\ref{momv}).

This procedure is more complex, and requires lengthier computations,
but we believe that it leads to a greater understanding of the
structure of the equations, and thus we present it here.  As an
additional benefit, this procedure provides explicit expressions for
$\dot{q}_A$ and $p^A$ in terms of each other, as well as explicit
expressions for the superhamiltonian and supermomenta.

It is convenient to rewrite equation (\ref{pvr}) for $p^A$ in
the form
\begin{equation} 
\label{pacn} 
p^A = \sqrt{g}\, \left( K g^{ij} {M^A}_{ij} - K^{ij} {M^A}_{ij}\right).
\end{equation} 
As before, we have 
\begin{equation} 
\label{kinq} 
K_{lm} = \frac{1}{2N}\, \left[ N_{l|m} + N_{m|l} - 
{M^B}_{lm} \dot{q}_B\right] 
\end{equation} 
which yields 
\begin{eqnarray} 
K &= & g^{lm}\, K_{lm} = \frac{1}{2N}\, 
\left[ g^{lm}\, (N_{l|m} + N_{m|l}) - g^{lm}\, {M^B}_{lm} \dot{q}_B\right] \\
K^{ij} & =& g^{il} g^{jm}\, K_{lm} = \frac{1}{2N}\, 
\left[ g^{il} g^{jm}\, (N_{l|m} + N_{m|l}) - 
g^{il} g^{jm}\, {M^B}_{lm} \dot{q}_B\right],
\end{eqnarray} 
and substituting the last two expressions into (\ref{pacn}) leads to
\begin{eqnarray} 
  p^A &=& \frac{\sqrt{g}}{2N}\, 
  \left[ -(g^{il} g^{jm} - g^{ij} g^{lm})\, {M^A}_{ij}\, 
    (N_{l|m} + N_{l|m})  \right. \nonumber \\ 
  & &\left. \quad \qquad + (g^{il} g^{jm} - g^{ij} g^{lm})\, {M^A}_{ij}\, {M^B}_{lm}\, 
    \dot{q}_B\right] .
\end{eqnarray} 
This can be written in a more compact form if we introduce the
notations
\begin{equation} 
G^{ij\, lm} = g^{il} g^{jm} - g^{ij} g^{lm}   
\end{equation} 
or its symmetrized version 
\begin{equation} 
G^{ij\, lm} =\frac{1}{2} g^{il} g^{jm} + \frac{1}{2} g^{jl} g^{im} 
- g^{ij} g^{lm}   
\end{equation} 
and 
\begin{equation} 
\label{qab} 
Q^{AB} = G^{ij\, lm}\, {M^A}_{ij}\, {M^B}_{lm} .
\end{equation} 
With these notations, the momenta becomes
\begin{equation} 
\label{pinq}
p^A = \frac{\sqrt{g}}{2N}\, 
\left[ -G^{ij\, lm}\, {M^A}_{ij}\, (N_{l|m} + N_{l|m}) + 
Q^{AB}\, \dot{q}_B\right]. 
\end{equation} 
We have assumed that the $q_A$ are independent.  This implies that the
square matrix $Q^{AB}$ (with dimension $n_q \times n_q$) has rank
$n_q$ and, consequently, is invertible.  In what follows we use the
same letters $G$ and $Q$ with lower indices for the elements of
the inverse matrices,
\begin{eqnarray} 
\left( G_{ij\, lm}\right) &=& \left( G^{ij\, lm}\right)^{-1} \\
\left( Q_{AB}\right) &=& \left( Q^{AB}\right)^{-1} ,
\end{eqnarray} 
so that 
\begin{equation} 
G_{ij\, kn}\, G^{kn\, lm} = \delta^l_i\, \delta^j_m 
\end{equation} 
and 
\begin{equation} 
Q_{AB}\, Q^{BC} = \delta^C_A .
\end{equation} 
These matrices have the quite obvious symmetry properties
\begin{equation} 
G^{ij\, lm} = G^{ji\, lm} = G^{ij\, ml} = G^{lm\, ij}
\end{equation}
\begin{equation} 
G_{ij\, lm} = G_{ji\, lm} = G_{ij\, ml} = G_{lm\, ij}
\end{equation}
as well as 
\begin{equation} 
Q^{AB} = Q^{BA};\qquad Q_{AB} = Q_{BA} .
\end{equation} 
There are two more useful relations that follow from simple observations. 
By definition
\begin{equation} 
Q_{BC}\, Q^{CA} = Q_{BC}\, {M^C}_{ij}\, G^{ij\, lm}\, {M^A}_{lm} = \delta^A_B 
\end{equation} 
which implies 
\begin{equation} 
{M^B}_{kn}\, Q_{BC}\, {M^C}_{ij}\, G^{ij\, lm}\, {M^A}_{lm} = 
\delta^A_B\, {M^B}_{kn} = {M^A}_{kn} = \delta^l_k\, \delta^m_n\, {M^A}_{lm} .
\end{equation} 
If $\{q_A\}^{n_q}_{A=1}$ are a complete set of variables ($g_{ik}$
depend only on $q_A$ and do not have any other arguments) then this
results in the two relations
\begin{equation} 
{M^B}_{kn}\, Q_{BC}\, {M^C}_{ij}\, G^{ij\, lm} = \delta^l_k\, \delta^m_n 
\end{equation} 
\begin{equation} 
{M^B}_{kn}\, Q_{BC}\, {M^C}_{ij} = G_{kn\, ij} 
\end{equation} 
that will be used, together with symmetries, in computations below. 

The expression (\ref{pinq}) for the momenta can be considered as a
system of linear equations for $\dot{q}_B$. Its solution 
\begin{equation} 
\label{qinp} 
\dot{q}_B = Q_{BA}\, G^{ij\, lm}\, {M^A}_{ij}\, (N_{l|m} + N_{m|l}) + 
2 N g^{-\frac{1}{2}}\, Q_{BA}\, p^A 
\end{equation} 
expresses $\dot{q}_B$ in terms of the momenta $p^A$. 

We begin to calculate the Hamiltonian $\widetilde\HH$ by expressing
the extrinsic curvature $K_{lm}$, $\TT (\KK )$ and $\TT (\KK^2)$ in
terms of momenta.  The last term in square brackets in equation
(\ref{kinq}) for $K_{lm}$ can be written in terms of momenta by
substituting $\dot{q}_B$ given by (\ref{qinp}):
\begin{eqnarray} 
{M^B}_{lm}\, \dot{q}_B &=& 
\underbrace{{M^B}_{lm}\, q_{BA}\, G^{ij\, lm}\, 
{M^A}_{ij}}_{\delta^k_l\, \delta^n_m}\, (N_{k|n} + N_{n|k})
   + 2 N g^{-\frac{1}{2}}\, Q_{BA}\, p^A\, {M^B}_{lm} \nonumber\\ 
 & = & N_{l|m} + N_{m|l} + 2 N g^{-\frac{1}{2}}\, Q_{BA}\, p^A\, {M^B}_{lm} .
\end{eqnarray} 
This results in
\begin{eqnarray} 
  K_{lm} &=& -g^{-\frac{1}{2}}\, Q_{BA}\, p^A\, {M^B}_{lm} \\
  \TT\KK &=& -g^{-\frac{1}{2}}\, Q_{BA}\, p^A\, g^{lm}\, {M^B}_{lm} \\
  K^{ij} & =& -g^{-\frac{1}{2}}\, Q_{BA}\, p^A\, g^{il}\, g^{jm}\, {M^B}_{lm} \\
  K_{ij} &=& -g^{-\frac{1}{2}}\, Q_{CD}\, p^D\, {M^C}_{ij} 
\end{eqnarray}
which, together with $\TT (\KK^2) = K^{ij}\, K_{ij}$, allows us to express 
the Lagrangian $\LL$ in terms of momenta:
\begin{eqnarray} 
\label{lgrp} 
\LL &=& \left[ R - (\TT\KK)^2 + \TT (\KK^2)\right]\, \sqrt{g}\, N
\nonumber \\ 
&=& N\, \left[ g^{\frac{1}{2}}\, R + g^{-\frac{1}{2}}\, Q_{BA}\, Q_{CD}\, 
\underbrace{G^{ij\, lm}\, {M^C}_{ij}\, {M^B}_{lm}}_{Q^{CB}}\, p^A\, 
p^D\right] \nonumber  \\ 
 &=& N\, \left[ g^{\frac{1}{2}}\, R + g^{-\frac{1}{2}}\, Q_{AB}\, p^A\, 
p^B\right]. 
\end{eqnarray} 
In addition, (\ref{qinp}) implies
\begin{equation} 
p^A\, \dot{q}_A = 2 N g^{-\frac{1}{2}}\, Q_{AB}\, p^A\, p^B + 
2\, N_{i|j}\, Q_{AB}\, p^A\, {M^B}_{lm}\, G^{lm\, ij} 
\end{equation} 
which results in the expression for the Hamiltonian 
\begin{eqnarray} 
  \widetilde\HH &=& p^A\, \dot{q}_A - \LL\ \hbox{\rm (Mod total
 divergence)}  \nonumber \\ 
 &=& N\, \left[ g^{-\frac{1}{2}}\, Q_{AB}\, p^A\, 
p^B - g^{\frac{1}{2}}\, R\right] + 2\, N_{i|j}\, Q_{AB}\, p^A\, {M^B}_{lm}\, 
G^{lm\, ij}   .
\end{eqnarray} 
The last term in this expression can be written as 
\begin{equation} 
\begin{array}{rl}  
2\, N_{i|j}\, Q_{AB}\, p^A\, {M^B}_{lm}\, G^{lm\, ij} &= 
\left( 2 N_i\, Q_{AB}\, p^A\, {M^B}_{lm}\, G^{lm\, ij}\right)_{|j}  \\ 
 & \\  
 &- N_i \left( 2\, Q_{AB}\, p^A\, {M^B}_{lm}\, G^{lm\, ij}\right)_{|j} 
\end{array} 
\end{equation} 
The covariant divergence of a vector density is a total divergence and 
can be thrown out. This reduces the Hamiltonian to 
\begin{equation} 
\label{hqorg} 
\widetilde\HH = N\, \left[ g^{-\frac{1}{2}}\, Q_{AB}\, p^A\, 
p^B - g^{\frac{1}{2}}\, R\right] + 
N_i \left( -2\, Q_{AB}\, p^A\, {M^B}_{lm}\, G^{lm\, ij}\right)_{|j} 
\end{equation} 
which is usually written as 
\begin{equation} 
\label{htt} 
\widetilde\HH = N\, \HH (q_A, p^A) + N_i\, \HH^i(q_A, p^A) 
\end{equation} 
where 
\begin{equation} 
\label{suph} 
\HH (q_A, p^A) = g^{-\frac{1}{2}}\, Q_{AB}\, p^A\, 
p^B - g^{\frac{1}{2}}\, R  
\end{equation} 
is the superhamiltonian, and 
\begin{equation} 
\label{supmm} 
\HH^i(q_A, p^A) = \left( -2\, Q_{AB}\, p^A\, {M^B}_{lm}\, 
G^{lm\, ij}\right)_{|j} 
\end{equation} 
are supermomenta in the superspace of $q_A$. It is trivial to verify,
by computing appropriate derivatives of (\ref{hqorg}), that the
Hamilton equation
\begin{equation} 
\label{triv} 
\dot{q}_A = \frac{\partial\widetilde\HH}{\partial p^A} 
\end{equation} 
simply reproduce the results obtained by inverting the definition  of
the momenta. 

The action can now be written in Hamiltonian form as
\begin{eqnarray} 
I_c &=& \int \left( p^A \dot{q}_A - 
\widetilde\HH (q_A, p^A, N, N_i)\right)\, d^3x\, dt \nonumber \\ 
 &=& \int \left( p^A \dot{q}_A - N\, \HH (q_A, p^A) - N_i\, 
\HH^i(q_A, p^A)\right)\, d^3x\, dt .
\end{eqnarray} 
Variation of this action with respect to $p^A$ and $q_A$ 
reproduce the Hamilton equations 
\begin{equation} 
\label{qham1}
\dot{q}_A = \frac{\partial\widetilde\HH}{\partial p^A} 
\end{equation} 
and 
\begin{equation} 
\dot{p}^A = -\frac{\partial\widetilde\HH}{\partial q_A} ,
\end{equation} 
while variations of shift and lapse yield the superhamiltonian and 
supermomenta constraint equations 
\begin{equation} 
\HH (q_A, p^A) = 0 
\end{equation} 
\begin{equation} 
\HH^i(q_A, p^A) = 0 
\end{equation}
that will be discussed in more detail after we develop convenient
notations and formal machinery to handle relevant questions. At this
point we only wish to stress once more that equations (\ref{qham1})
can be obtained by inverting the kinematic relations (\ref{pinq}) and,
thus, can be treated as independent of the Hamiltonian dynamics.

\section{Geometrodynamic Superspace.} 
\label{IV} 

York's analysis of the geometrodynamic degrees of freedom suggests
that the appropriate configuration space for geometrodynamics is not
the superspace of 3--metrics (or 3--geometries), but rather the space
of conformal 3--geometries.  We describe here the ideas of such
dynamics in a generalized form for a case when the 3-metric components
$g_{ik}$ are given as functions of $n_q$ other variables $q_A$, $A = 1,
\ldots , n_q \le 6$,
\begin{equation} 
g_{ij} = g_{ij}(q_A). 
\end{equation} 
The functions $q_A$ are assumed to be independent and form a complete
set.  Following York's analysis, we split the set of variables
$\{q_A\}_{A=1}^{n_q}$ into a subset $\{\beta_I\}_{I=1}^{n_d}$ ($n_d
\le 2$) of the true dynamic variables and a subset
$\{\alpha_\mu\}_{\mu = 0}^{n_e}$ ($n_e \le 3$) of the so-called
embedding variables, with the identifications facilitating a
comparison with the equations of the previous section:
\begin{equation} 
\beta_I = q_I 
\end{equation} 
\begin{equation} 
\alpha_\mu = q_{n_d + \mu + 1} \equiv q_\mu  .
\end{equation} 
This allows us to freely switch notations to better suit the context.
It is clear that
\begin{equation} 
n_q = n_d + n_e + 1 .
\end{equation} 

We wish to reformulate geometrodynamics on the configuration space
(geometrodynamic superspace) of the true dynamical degrees of freedom
$q_I = \beta_I$ (conformal superspace as opposed to the superspace of
3--metrics of the previous two sections).

The Lagrangian $\LL$, equation (\ref{lagr}) with $\KK$ given by
(\ref{kkq}), remains a function of the same variables as in the
previous section.  It can be written more appropriately as
\begin{eqnarray}
\LL &=& \LL (q_A, \dot{q}_A, N, N_i) \nonumber \\
 &=& \LL (q_I, \dot{q}_I, q_\mu , 
\dot{q}_\mu ,N, N_i) = \LL (\beta_I, \dot{\beta}_I, {\alpha}_\mu , 
\dot{\alpha}_\mu ,N, N_i),
\end{eqnarray} 
where only $q_I$, $(\dot q)_I$ are related to the new configuration
space.  The rest of the arguments of $\LL$ are functional parameters,
which make the Lagrangian $\LL$ explicitly time dependent (although
the time dependence is introduced through functions $q_\mu ,
\dot{q}_\mu ,N, N_i$).  Only the momenta $p^I$ conjugate to $q_I$
\begin{equation} 
p^I = \frac{\partial\LL}{\partial\dot{q}_I} 
\end{equation} 
retain their dynamical meaning (and eventually become arguments of the
Hamiltonian).  The similar quantities
\begin{equation} 
p^\mu = \frac{\partial\LL}{\partial\dot{q}_\mu} 
\end{equation} 
can be introduced and used in developing the theory, but cannot be treated  
as momenta. 

The transition to a Hamiltonian formulation, and the analysis of the
constraints, can be performed almost the same as in section \ref{III}.
It is especially simple in the case when the matrix $Q^{AB}$, given by
(\ref{qab}), is such that $Q^{\mu I} = Q^{I\mu} = 0$.  That is, $Q^{AB}$
has the block structure
\begin{equation} 
\left( Q^{AB}\right) = 
\left(\begin{array}{cc} Q^{IJ} & 0 \\ 0 & Q^{\mu\nu} \end{array} \right)
\end{equation} 
in which case the inverse matrix $Q_{AB}$ has the structure 
\begin{equation} 
\left( Q_{AB}\right) = 
\left(\begin{array}{cc} Q_{IJ} & 0 \\ 0 & Q_{\mu\nu} \end{array} \right)
\end{equation} 
where $Q_{IJ}$ is the inverse matrix of $Q^{IJ}$ and $Q_{\mu\nu}$ is
the inverse matrix of $Q^{\mu\nu}$.  This case is rather common in
applications (for instance, it includes all diagonal homogeneous
cosmologies).  The true meaning is determined by the expressions for
$p^I$ and $p^\mu$, replacing (\ref{pinq}) with
\begin{equation} 
\label{pinqI}
p^I = \frac{\sqrt{g}}{2N}\, 
\left[ -G^{ij\, lm}\, {M^I}_{ij}\, (N_{l|m} + N_{l|m}) + 
Q^{IJ}\, \dot{q}_J\right] 
\end{equation} 
and 
\begin{equation} 
\label{pinqmu}
p^\mu = \frac{\sqrt{g}}{2N}\, 
\left[ -G^{ij\, lm}\, {M^\mu}_{ij}\, (N_{l|m} + N_{l|m}) + 
Q^{\mu\nu}\, \dot{q}_\nu\right] .
\end{equation} 
It is important to notice that the $p^\mu$ depend only on time
derivatives of $q_\mu$, and do not involve time derivatives of the
true dynamical variables.

All the computations in the previous section can be repeated
literally, up until the Lagrangian is expressed in terms of momenta,
equation (\ref{lgrp}).  In view of (\ref{pinqI}) and (\ref{pinqmu}),
the final expression for $\LL$ takes the form
\begin{equation} 
\label{lgrpd} 
\begin{array}{rl} 
\LL &= N\, \left[ g^{\frac{1}{2}}\, R + g^{-\frac{1}{2}}\, Q_{AB}\, 
p^A\, p^B \right]  \\  
 & \\ 
 &= N\, \left[ g^{\frac{1}{2}}\, R + g^{-\frac{1}{2}}\, \left( Q_{IJ}\, 
p^I\, p^J + Q_{\mu\nu}\, p^\mu\, p^\nu\right)\right] ,
\end{array} 
\end{equation} 
which is essentially the same as in the previous section, except now
the $p^\mu$ are not momenta but merely functions
\begin{equation} 
p^\mu = p^\mu (q_A, \dot{q}_\mu , N, N_i) 
\end{equation} 
given by (\ref{pinqmu}).  This makes the Lagrangian $\LL$ a function
of the new arguments
\begin{equation} 
\LL = \LL (q_A, p^I, \dot{q}_\mu , N, N_i) = 
\LL (q_I, q_\mu , p^I, \dot{q}_\mu , N, N_i) .
\end{equation} 

The Hamiltonian $\HH_{DYN}$ on the dynamical superspace 
\begin{equation} 
  \label{hdyn}  
  \HH_{DYN} = p^I \dot{q}_I - \LL (q_A, p^I, \dot{q}_\mu , N, N_i) ,
\end{equation} 
is distinctly different from the Hamiltonian $\widetilde\HH$ on the
superspace of 3-metrics described in the previous section.  A useful
form of this Hamiltonian can be obtained by writing it as
\begin{equation} 
\label{hdync} 
\HH_{DYN} = \underbrace{p^A \dot{q}_A - 
  \LL (q_A, p^I, \dot{q}_\mu , N, 
  N_i)}_{\widetilde\HH (q_A, p^I, \dot{q}_\mu , N, N_i)} 
- p^\mu \dot{q}_\mu ,
\end{equation} 
where the first two terms form $\widetilde\HH$ of the previous
section, given by equations (\ref{htt})--(\ref{supmm}), with
expressions (\ref{pinqmu}) substituted for $p^\mu$.  More precisely,
\begin{equation} 
\label{hdynf} 
\HH_{DYN} = \widetilde\HH (q_A, p^I, \dot{q}_\mu , N, N_i)  
- p^\mu \dot{q}_\mu 
\end{equation} 
where 
\begin{equation} 
\label{httd} 
\widetilde\HH (q_A, p^I, \dot{q}_\mu , N, N_i) = N\, 
\HH (q_A, p^I, \dot{q}_\mu , N, N_i) + N_i\, \HH^i(q_A, p^I, \dot{q}_\mu , 
N, N_i)  
\end{equation} 
with   
\begin{equation} 
\label{suphd} 
\begin{array}{rl} 
\HH (q_A, p^I, \dot{q}_\mu , N, N_i) &= g^{-\frac{1}{2}}\, Q_{AB}\, p^A\, 
p^B - g^{\frac{1}{2}}\, R  \\ 
 & \\ 
 &= g^{-\frac{1}{2}}\, \left( Q_{IJ}\, p^I\, p^J + Q_{\mu\nu}\, p^\mu\, p^\nu 
\right)  
- g^{\frac{1}{2}}\, R  
\end{array} 
\end{equation} 
and  
\begin{eqnarray} 
\label{supmmd} 
\HH^i(q_A, p^I, \dot{q}_\mu , N, N_i) &=& \left( -2\, Q_{AB}\, p^A\, 
{M^B}_{lm}\, G^{lm\, ij}\right)_{|j} \nonumber \\
 &=& \left[ -2\, \left( Q_{IJ}\, p^I\, 
{M^J}_{lm} + Q_{\mu\nu}\, p^\mu\, {M^\nu}_{lm}\right)\, 
G^{lm\, ij}\right]_{|j} .
\end{eqnarray} 
The Hamiltonian action on the geometrodynamic superspace,
\begin{eqnarray} 
I &=& \int \left[ p^I \dot{q}_I - \HH_{DYN}\right]\, d^3x\, dt
 \nonumber \\ 
 &=& \int \left[ p^I \dot{q}_I - \left(\widetilde\HH (q_A, p^I, \dot{q}_\mu , 
N, N_i) - p^\mu \dot{q}_\mu\right)\right]\, d^3x\, dt 
\end{eqnarray} 
can be used to derive $n_d$ pairs of Hamilton equations 
\begin{eqnarray} 
\label{hameq} 
 \dot{q}_I &=& \frac{\partial\HH_{DYN}}{\partial p^I} \nonumber \\ 
 \dot{p}^I &=& \frac{\partial\HH_{DYN}}{\partial q_I} 
\end{eqnarray} 
by varying $p^I$ and $q_I$. 

Variations of the action with respect to $N$ and $N_i$ yield $n_c$
constraints, with $n_c \le 4$.  In view of equations (\ref{hdynf}) --
(\ref{supmmd}), the functional dependence of $\HH_{DYN}$ on $N$, $N_i$
can be expressed as 
\begin{equation} 
\begin{array}{rl} 
\label{hhnn} 
 &\HH_{DYN}\left( q_A, p^I, p^\mu (N, N_i, \dot{q}_\mu ), N, N_i\right)  \\
 & \\ 
 &= \underbrace{N\, \HH\left( q_A, p^I, p^\mu (N, N_i, 
\dot{q}_\mu )\right) + N_i\, \HH^i\left( q_A, p^I, p^\mu (N, N_i, 
\dot{q}_\mu )\right)}_{\widetilde\HH\left( q_A, p^I, p^\mu (N, N_i, 
\dot{q}_\mu ), N, N_i\right)}  \\
&\\
&\qquad \qquad -\, p^\mu (N, N_i, \dot{q}_\mu )\, \dot{q}_\mu .
\end{array} 
\end{equation} 
Variation of $N$ produces the constraint 
\begin{equation} 
\label{shdtt} 
\frac{\partial\HH_{DYN}}{\partial N} + 
\frac{\partial\HH_{DYN}}{\partial p^\mu}\, \frac{\partial p^\mu}{\partial N} 
= 0 ,
\end{equation} 
and equation (\ref{hhnn}) implies 
\begin{equation} 
\frac{\partial\HH_{DYN}}{\partial N} = 
\HH\left( q_A, p^I, p^\mu (N, N_i, \dot{q}_\mu )\right),
\end{equation} 
so, together with the trivial non-dynamic expression (\ref{triv}) of the 
previous section,
\begin{equation} 
\frac{\partial\HH_{DYN}}{\partial p^\mu} = 
\frac{\partial\widetilde\HH}{\partial p^\mu} - \dot{q}_\mu = 
 \dot{q}_\mu - \dot{q}_\mu = 0,
\end{equation} 
which reduces (\ref{shdtt}) to the superhamiltonian constraint 
\begin{equation} 
\HH\left( q_A, p^I, p^\mu (N, N_i, \dot{q}_\mu )\right) = 0 .
\end{equation} 
Likewise, variation of $N_i$ leads to the supermomenta constraints 
\begin{equation} 
\HH^i\left( q_A, p^I, p^\mu (N, N_i, \dot{q}_\mu )\right) = 0 .
\end{equation} 
These superhamiltonian and supermomenta constraints on the geometrodynamic 
superspace are obtained from the constraints on the superspace of 
3--geometries by a simple substitution of $p^\mu$ as given by 
(\ref{pinqmu}). 

In the general case the matrix $Q^{AB}$ does not have a block
structure.  Instead, it can be written as
\begin{equation} 
\left( Q^{AB}\right) = 
\left(\begin{array}{cc} Q^{IJ} & Q^{I\nu} \\ 
Q^{\mu J} & Q^{\mu\nu} \end{array} \right) ,
\end{equation} 
in which case the inverse matrix $\widetilde{Q}_{AB}$ has the structure 
\begin{equation} 
\left( \widetilde{Q}_{AB}\right) = 
\left(\begin{array}{cc} \widetilde{Q}_{IJ} & \widetilde{Q}_{I\nu} \\ 
\widetilde{Q}_{\mu J} & \widetilde{Q}_{\mu\nu} \end{array} \right) .
\end{equation} 
It is clear that $\widetilde{Q}_{IJ} \not= Q_{IJ}$ and that
$\widetilde{Q}_{IJ}$ is not the inverse matrix of the sub-matrix
$Q^{IJ}$.  Likewise, $\widetilde{Q}_{\mu\nu} \not= Q_{\mu\nu}$ and
$\widetilde{Q}_{\mu\nu}$ is not the inverse matrix of the sub-matrix
$Q^{\mu\nu}$.

Expressions for $p^I$ and $p^\mu$ take the form (compare with (\ref{pinqI}) and 
(\ref{pinqmu})):
\begin{equation} 
\label{pinqIg}
p^I = \frac{\sqrt{g}}{2N}\, 
\left[ -G^{ij\, lm}\, {M^I}_{ij}\, (N_{l|m} + N_{l|m}) + 
Q^{IJ}\, \dot{q}_J + Q^{I\nu}\, \dot{q}_\nu\right] 
\end{equation} 
and 
\begin{equation} 
\label{pinqmug}
p^\mu = \frac{\sqrt{g}}{2N}\, 
\left[ -G^{ij\, lm}\, {M^\mu}_{ij}\, (N_{l|m} + N_{l|m}) + 
Q^{\mu J}\, \dot{q}_J + Q^{\mu\nu}\, \dot{q}_\nu\right] .
\end{equation} 
An important change is that now the $p^\mu$ depend not only on time
derivatives of $q_\mu$ but also on time derivatives of the true
dynamical variables.

Expressing the Lagrangian $\LL$ in terms of momenta yields
\begin{equation} 
\label{lgrpdg}  
\LL = N\, \left[ g^{\frac{1}{2}}\, R + g^{-\frac{1}{2}}\, 
\left( \widetilde{Q}_{IJ}\, p^I\, p^J + \widetilde{Q}_{I\nu}\, p^I\, p^\nu + 
\widetilde{Q}_{\mu J}\, p^\mu\, p^J + 
\widetilde{Q}_{\mu\nu}\, p^\mu\, p^\nu\right)\right] 
\end{equation} 
which is again essentially the same as in the previous section, except
now $p^\mu$ are functions
\begin{equation} 
\label{inm}
p^\mu = p^\mu (q_A, \dot{q}_I, \dot{q}_\mu , N, N_i) 
\end{equation} 
given by (\ref{pinqmug}).  This in turn makes the Lagrangian $\LL$ a
function of the new arguments
\begin{equation} 
\LL = \LL (q_A, p^I, p^\mu (q_A, \dot{q}_I, \dot{q}_\mu , N, N_i), N, N_i) .
\end{equation} 
The dependence of the Lagrangian on $\dot{q}_I$ in the general case
seems to break the argument developed above in the case when $Q^{AB}$
has block structure. In fact, this is not the case, and the problem
can be easily remedied.

As before, we assume that the true dynamical variables $q_I$ are
independent, which implies that the sub-matrix $Q^{IJ}$ in
(\ref{pinqIg}) is invertible (we keep the notation $Q_{IJ}$ for the
elements of the matrix inverse to the sub-matrix $Q^{IJ}$).  This means
that (\ref{pinqIg}) can be considered as a system of linear equations
with respect to $\dot{q}_I$.   Its solution
\begin{equation} 
\label{qupgg} 
\dot{q}_I = Q_{IJ}\, G^{ij\, lm}\, {M^J}_{ij}\, (N_{l|m} - N_{l|m}) - 
Q_{IJ}\, Q^{J\nu}\, \dot{q}_\nu + 2 N g^{-\frac{1}{2}}\, Q_{IJ}\, p^J   
\end{equation} 
expresses $\dot{q}_I$ as functions 
\begin{equation} 
\label{notqd} 
\dot{q}_I = \dot{q}_I(q_A, p^I, \dot{q}_\mu, N, N_i) .
\end{equation} 
Their substitution in (\ref{inm}) transforms $p^\mu$ to
\begin{equation} 
\label{notpd} 
p^\mu = p^\mu \left( q_A, \dot{q}_I(q_A, p^I, \dot{q}_\mu, N, N_i), 
\dot{q}_\mu , N, N_i\right) = p^\mu(q_A, p^I, \dot{q}_\mu , N, N_i) 
\end{equation} 
and $\LL$ into the form
\begin{equation} 
\begin{array}{rl} 
\LL &= \LL \left( q_A, p^I, p^\mu \left( q_A, 
\dot{q}_I(q_A, p^I, \dot{q}_\mu, N, N_i), 
\dot{q}_\mu , N, N_i\right) , N, N_i\right) = \\ 
 & \\
 &= \LL (q_A, p^I, \dot{q}_\mu , N, N_i) .
\end{array} 
\end{equation} 
With this in mind, we can follow the same steps as before in
developing the Hamiltonian formalism. We introduce the geometrodynamic
Hamiltonian as
\begin{equation} 
\HH_{DYN} = p^I \dot{q}_I - \LL 
\end{equation} 
where $\LL$ is given by (\ref{lgrpdg}). This allows us to recover the
same expressions for $\HH_{DYN}$
\begin{equation} 
\HH_{DYN} = p^A \dot{q}_A - \LL - p^\mu \dot{q}_\mu = 
\widetilde\HH - p^\mu \dot{q}_\mu ,
\end{equation} 
with 
\begin{equation} 
\widetilde\HH = N\, \HH + N_i\, \HH^i 
\end{equation} 
where $\widetilde\HH$, $\HH$, $\HH^i$ are given by the same
expressions as in the previous section, except that $p^\mu$ is now
given by (\ref{pinqmug}) and $\dot{q}_I$ is given by (\ref{qupgg})
(also, $Q_{AB}$ from the previous section should be replaced by
$\widetilde{Q}_{AB}$).

Tracing the chain of arguments developed above results in the action
\begin{eqnarray} 
I &=& \int \left[ p^I \dot{q}_I - \HH_{DYN}\right]\, d^3x\, dt \nonumber \\ 
 &=& \int \left[ p^I \dot{q}_I - \left(\widetilde\HH - 
p^\mu \dot{q}_\mu\right)\right]\, d^3x\, dt ,
\end{eqnarray} 
with $\HH_{DYN}$ and $\widetilde\HH$ arguments best described by equations 
\begin{eqnarray} 
\label{arghd} 
\HH_{DYN} &=& \HH_{DYN} \left( q_A, p^I, p^\mu \left( q_A, 
    \dot{q}_I, \dot{q}_\mu , N, N_i\right) , N, N_i\right) \nonumber\\
\widetilde\HH &=& \widetilde\HH \left( q_A, p^I, p^\mu \left( q_A, 
    \dot{q}_I, \dot{q}_\mu , N, N_i\right) , N, N_i\right) \\ 
\dot{q}_I &=& \dot{q}_I(q_A, p^I, \dot{q}_\mu, N, N_i)\nonumber .
\end{eqnarray} 
Following the same line of arguments as before leads to the Hamilton
equations (\ref{hameq}) and to the $n_c$ constraint equations
\begin{eqnarray} 
\HH\left( q_A, p^I, p^\mu (N, N_i, \dot{q}_I(q_a, p^I, \dot{q}_\mu , N, N_i), 
\dot{q}_\mu )\right) &=& 0 \\
\HH^i\left( q_A, p^I, p^\mu (N, N_i, 
\dot{q}_I(q_a, p^I, \dot{q}_\mu , N, N_i), \dot{q}_\mu )\right) &= & 0 .
\end{eqnarray} 
These superhamiltonian and supermomenta constraints on the geometrodynamic 
superspace in the general case are obtained from the constraints on the superspace of 
3--geometries by a simple substitution of $p^\mu$ as given by 
(\ref{pinqmug}) and $\dot{q}_I$ as given by (\ref{qupgg}).

\section{The Hamilton--Jacobi Equation.} 
\label{V} 

A detailed description of the Hamilton--Jacobi equation on the
configuration space of 3--metrics $g_{ik}$ can be found in the
literature, including monographs, together with a detailed explanation
of how the concept of a functional (bubble) derivative takes an active
part in writing the equation. We therefore provide only a very brief
discussion of this subject here.

Following standard approaches we introduce the Hamilton principal
functional,
\begin{equation} 
\label{Stg3} 
S = S[t, g_{ik}] 
\end{equation} 
as the extremal value of the action (\ref{icosm}), evaluated between
the 3-slices given by $(t', g'_{ik})$ and $(t, g_{ik})$.  The primed
slice is assumed to be fixed:
\begin{equation} 
\label{phfunc}
S[t, g_{ik}] = I_{extremum} = \int\limits_{(t', g'_{ik})}^{(t, g_{ik})} 
 \left(\pi^{ij} \dot{g}_{ij} - \HH_{WDW}\right)\, d^3x\, dt 
\end{equation} 
The integral on the right hand side of this expression is extremized
with both ends fixed. If the upper limit is released, the integral
becomes a functional of $t$ and $g_{ik}$ given by (\ref{phfunc}). The
Hamilton--Jacobi equation for this functional is obtained by variation
of the upper limit.  Sometimes this variation is preceded by imposing
the constraints\cite{MTW70}.  However, such a move weakens arguments
following the variational procedure.  Instead one can vary the
integral before imposing the constraints, as it is done in mechanics.  It
is easy to see that the variation $(\delta t, \delta g_{ik})$ on the
final hypersurface produces variation in $S$ given by
\begin{equation} 
\delta S = \int \left[\pi^{ik}\, \delta g_{ik} - \HH_{WDW}(g_{ik}, 
\pi^{ik})\, \delta t\right]\, d^3x .
\end{equation} 
The standard expression for this variation in terms of functional derivatives 
of $S$ is 
\begin{equation} 
\delta S = \int \left[\frac{\delta S}{\delta t}\, \delta t + 
\frac{\delta S}{\delta g_{ik}}\, \delta g_{ik}\right]\, d^3x ,
\end{equation} 
and comparison of the two equations yields the expression for momenta
\begin{equation} 
\pi^{ik} = \frac{\delta S}{\delta g_{ik}} 
\end{equation} 
and one more equation 
\begin{equation}    
\label{hjsmch}
\frac{\delta S}{\delta t} = - \HH_{WDW}(g_{ik}, \pi^{ik}).
\end{equation} 
Together, these result in a functional differential equation for $S$,
\begin{equation} 
\label{hjeqg} 
-\frac{\delta S}{\delta t} = \HH_{WDW}\left( g_{ik}, 
\frac{\delta S}{\delta g_{ik}}\right) 
\end{equation} 
that should be considered the Hamilton--Jacobi equation.  However, in
Hamiltonian geometrodynamics on the superspace of 3--metrics, the
constraints (\ref{hcns}) and (\ref{hicns}) change the nature of this
equation in such a way that, although being an important statement
concerning the nature of time, it no longer performs the same
functions as the Hamilton--Jacobi equation in mechanics, at least in
the standard treatment of the subject.  The standard approach is to
first impose the constraints, which take the form
\begin{eqnarray} 
\label{hscns}
\HH\left( g_{ij}, \frac{\delta S}{\delta g_{ik}}\right) &=& 0 \\
\label{hiscns}
\HH^i\left( g_{ij}, \frac{\delta S}{\delta g_{ik}}\right) &=& 0 .
\end{eqnarray}   
Equation (\ref{hamg}) for $\HH_{WDW}$ then allows us to write the
Hamilton--Jacobi equation (\ref{hjeqg}) in the form
\begin{equation} 
\label{hjeqgg} 
-\frac{\delta S}{\delta t} = N\,\HH\left( g_{ik}, 
\frac{\delta S}{\delta g_{ik}}\right) + N_i\,\HH^i\left( g_{ik}, 
\frac{\delta S}{\delta g_{ik}}\right)
\end{equation} 
which, together with (\ref{hscns}) and (\ref{hiscns}) yields 
\begin{equation} 
\label{der0} 
\frac{\delta S}{\delta t} = 0 .
\end{equation} 
From the dynamical point of view the partial functional derivative of
$S$ with respect to $t$ is computed with only $g_{ik}$ fixed. Equation
(\ref{der0}) might seem strange, since (\ref{phfunc}) implies that $S$
also depends on $N$ and $N_i$, which might in turn depend on $t$.
This does not create problems because, as is easy to see,
\begin{equation} 
\frac{\delta S}{\delta N} = \frac{\delta S}{\delta N_i} = 0 .
\end{equation} 
This means that equation (\ref{der0}) only states that $S$ does not
depend explicitly on time and that any dependence on $t$ can emerge
only through the components of $g_{ik}$.  In other words, information
about time is carried by the 3--metric of the slice, which, after
proper refinements provides the basis for Baierlane, Sharp and
Wheeler's concept of intrinsic time.  It also implies that the
functional $S[t, g_{ik}]$ is a functional of the slice 3--metric only
\begin{equation} 
\label{Sg3}
S = S[g_{ik}] .
\end{equation} 

In any case, equation (\ref{hjsmch}), although reminiscent of the
Hamilton--Jacobi equation of mechanics, does not perform functions
that are expected from the Hamilton--Jacobi equation.  This statement 
requires
more refined considerations, based on the observation that equations
(\ref{hiscns}) can be written as (cf. (\ref{supmg}))
\begin{equation} 
\label{dinv} 
\left(\frac{\delta S}{\delta g_{ij}}\right)_{|j} = 0,  
\end{equation} 
which can be interpreted\cite{MTW70} as invariance of $S$ with 
respect to the choice of coordinates (diffeomorphism invariance), and 
can be expressed by the statement that $S$ is not even the functional of 
the 3--metric, but only of its diffeomorphically invariant part, called 
the 3--geometry 
${}^{(3)}\GG$ 
\begin{equation} 
\label{SG3} 
S = S[{}^{(3)}\GG] .
\end{equation}  
This reduces the left hand side of equation (\ref{hjeqgg}), together
with the second term on the right hand side of the same equation, to
the kinematic statement expressed by (\ref{SG3}).  The only remaining
part that can possibly have dynamic content can be written as
\begin{equation}
\label{HJG} 
\HH\left({}^{(3)}\GG , \frac{\delta S}{\delta {}^{(3)}\GG}\right) = 0 .
\end{equation} 
This equation is identified as the Hamilton--Jacobi equation on the
superspace of 3--geometries.  It can be shown to encode all the
dynamic content of the theory, but the equation is purely symbolic. In
practice, this equation can be solved by picking for $S[g_{ik}]$ a
form satisfying (\ref{dinv}), and subsequently adjusting the
functional parameters of this solution to satisfy (\ref{hscns}).
Alternatively, we could in principle solve equation (\ref{hjsmch}) or
(\ref{hjeqgg}) and then adjust parameters of the solution to satisfy
(\ref{hscns}) and (\ref{hiscns}).  The first step of this procedure
produces solutions of (\ref{hjsmch}) both on and off the constraint
shell, which involves providing information (including but not limited
to information concerning shift and lapse) that gets eliminated when
the solutions are restricted to the shell. This procedure is
equivalent to the first one, but is not as practical in classical
geometrodynamics.  We will return to it later in the context of
quantum geometrodynamics.

The transformation of variables described in section \ref{III}, and
given by equations (\ref{vtr}), result from replacing the variables
$g_{ik}$ with the new variables $q_A$.  This modifies the equations
but leaves intact the content of the theory and its interpretation in
the generic case.  Slight modifications are required in non-generic
cases, when the dimension of the configuration superspace is reduced.

On the superspace of $q_A$, the Hamilton principal functional
\begin{equation} 
\label{Stq3} 
S = S[t, q_A] 
\end{equation} 
is
\begin{equation} 
\label{phfunq}
S[t, q_A] = I_{extremum} = \int\limits_{(t', {}q'_A)}^{(t, q_A)} 
 \left(p^A \dot{q}_A - \widetilde\HH (N, N_i, q_A, p^A)\right)\, d^3x\, dt 
\end{equation}  
where $\widetilde\HH$ is given by equations (\ref{hqorg}) --
(\ref{supmm}).  The variational procedure on the superspace of $q_A$
is similar to the one used on the superspace of $g_{ik}$, and yields
the expression for the momenta
\begin{equation} 
p^A = \frac{\delta S}{\delta q_A} 
\end{equation} 
and the equation
\begin{equation}    
\label{hjsmchq}
\frac{\delta S}{\delta t} = - \widetilde\HH (N, N_i, q_A, p^A).
\end{equation} 
Together these result in the functional differential equation for $S$,
\begin{equation} 
\label{hjeqq} 
-\frac{\delta S}{\delta t} = \widetilde\HH\left( N, N_I, q_A, 
\frac{\delta S}{\delta q_A}\right).
\end{equation} 
Following the logic of standard Hamiltonian dynamics this is
considered to be the Hamilton--Jacobi equation. In view of
(\ref{htt}), this equation can be written as
\begin{equation} 
\label{hjeqgq} 
-\frac{\delta S}{\delta t} = N\,\HH\left( q_A, 
\frac{\delta S}{\delta q_A}\right) + N_i\,\HH^i\left( q_A, 
\frac{\delta S}{\delta q_A}\right).
\end{equation} 
The same line of reasoning as before leads to the constraints
\begin{eqnarray} 
\label{hscnsq}
\HH\left( q_A, \frac{\delta S}{\delta q_A}\right) &=& 0 \\
\label{hiscnsq}
\HH^i\left( q_A, \frac{\delta S}{\delta q_A}\right) &=& 0 ,
\end{eqnarray} 
which can also be expressed as
\begin{equation} 
\frac{\delta S}{\delta N} = \frac{\delta S}{\delta N_i} = 0 .
\end{equation} 
The three equations (\ref{hjeqgq})--(\ref{hiscnsq}) imply, as before,
\begin{equation} 
\label{STG} 
\frac{\delta S}{\delta t} = 0 .
\end{equation} 
The last two equations can be summarized by the statement that
$S[t,q_A]$ is, in fact, a functional of $q_A$ only:
\begin{equation} 
\label{Sgq3}
S = S[q_A] .
\end{equation}  

Additional considerations, similar to those leading to equations
(\ref{dinv}) -- (\ref{HJG}), are based on the observation that
equations (\ref{hiscnsq}) can be written as (cf. (\ref{supmm}))
\begin{equation} 
\label{dinvq} 
\left( -2\, Q_{AB}\, \frac{\delta S}{\delta q_A}\, {M^B}_{lm}\, 
G^{lm\, ij}\right)_{|j} = 0    
\end{equation} 
which can again be interpreted, in the generic case, as invariance of
$S$ with respect to the choice of coordinates (diffeomorphism
invariance).  This can be expressed by the statement that $S$ is not
even a functional of $q_A$, but only of the diffeomorphically invariant 
information
carried by $q_A$, the 3--geometry ${}^{(3)}\GG$ (cf. equation
(\ref{SG3})).  This reduces the left hand side of equation
(\ref{hjeqgq}), together with the second term on the right hand side
of the same equation, to the kinematic statement expressed by
(\ref{SG3}).  The only remaining part that can possibly have dynamic
content can be expressed by (\ref{HJG}), or in practice, by equation
(\ref{hscnsq}), which might be identified as the Hamilton--Jacobi
equation on the configuration superspace of $q_A$.  Using
(\ref{suph}), we can also write
\begin{equation} 
\label{HJGq} 
g^{-\frac{1}{2}}\, Q_{AB}\, \frac{\delta S}{\delta q_A}\, 
\frac{\delta S}{\delta q_B} - g^{\frac{1}{2}}\, R = 0   .
\end{equation} 
All the comments made above concerning the solution of the
Hamilton--Jacobi equation on the superspace of $g_{ik}$ can be
repeated with no essential change in the generic case of the
superspace of $q_A$.

The non-generic, degenerate, cases require more attention.  A
straightforward application of variational principles in these cases
might not lead to the desired result because of restrictions imposed
on variations of the action.  These arise from the fixed form of
expressions for $g_{ik}$, as well as restrictions on variations of the
shift and lapse (often referred to as gauge conditions).

The geometrodynamic superspace has been described in section \ref{IV}.
It can be thought of as the configuration space of the true dynamical
variables $q_I$. The Hamilton principal functional on the
geometrodynamic superspace
\begin{equation} 
S = S[t, q_I] 
\end{equation} 
is given by 
\begin{equation} 
\label{phfunD}
S[t, q_I] = I_{extremum} = \int\limits_{(t', {}q'_I)}^{(t, q_I)} 
 \left(p^I \dot{q}_I - \HH_{DYN}\right)\, d^3x\, dt 
\end{equation}  
where 
\begin{equation} 
\HH_{DYN} = \widetilde\HH - p^\mu \dot{q}_\mu ,
\end{equation} 
and as before,
\begin{equation} 
\widetilde\HH = N\, \HH + N_i\, \HH^i .
\end{equation}  
The arguments of $\HH_{DYN}$ and $\widetilde\HH$ are given by
(\ref{arghd}), with functions $\dot{q}_I$ and $p^\mu$ given by
equations (\ref{notqd}) and (\ref{notpd}). In the discussion of the
Hamilton--Jacobi equation below, it is useful to describe the
arguments of $\HH_{DYN}$, $\widetilde\HH$, $\HH$, and $\HH^i$ as
follows
\begin{eqnarray} 
\label{arghD} 
\HH_{DYN} &=& 
 \HH_{DYN} \left( q_A, p^I, p^\mu \left( q_A, 
\dot{q}_I(q_A, p^I, \dot{q}_\mu, N, N_i), \dot{q}_\mu , N, N_i\right) , N, 
N_i\right) \nonumber \\
\widetilde\HH &=& \widetilde\HH \left( q_A, p^I, p^\mu \left( q_A, 
\dot{q}_I(q_A, p^I, \dot{q}_\mu, N, N_i), \dot{q}_\mu , N, N_i\right) , N, 
N_i\right) \\ 
\HH &=& \HH \left( q_A, p^I, p^\mu \left( q_A, 
\dot{q}_I(q_A, p^I, \dot{q}_\mu, N, N_i), \dot{q}_\mu , N, N_i\right) , N, 
N_i\right) \nonumber \\ 
\HH^i &=& \HH^i \left( q_A, p^I, p^\mu \left( q_A, 
\dot{q}_I(q_A, p^I, \dot{q}_\mu, N, N_i), \dot{q}_\mu , N, N_i\right) , N, 
N_i\right)  \nonumber.
\end{eqnarray} 
The variational procedure on the geometrodynamic superspace of $q_I$ ,
similar to the one used before (variation of the endpoints), yields
an expression for the momenta conjugate to the true dynamic variables
$q_I$,
\begin{equation} 
p^I = \frac{\delta S}{\delta q_I} ,
\end{equation} 
and the equation 
\begin{equation}    
\label{hjsmchD}
\frac{\delta S}{\delta t} = - \HH_{DYN} \left( q_A, p^I, p^\mu \left( q_A, 
\dot{q}_I(q_A, p^I, \dot{q}_\mu, N, N_i), \dot{q}_\mu , N, N_i\right) , N, 
N_i\right),
\end{equation} 
which together result in the functional differential equation for $S$
\begin{equation} 
\label{hjeqD} 
-\frac{\delta S}{\delta t}  
= \HH_{DYN} \left( q_A, 
\frac{\delta S}{\delta q_I}, p^\mu \left( q_A, 
\dot{q}_I\left( q_A, \frac{\delta S}{\delta q_I}, \dot{q}_\mu, 
N, N_i\right) , \dot{q}_\mu , N, N_i\right) , N, N_i\right)
\end{equation} 
This is considered to be the Hamilton--Jacobi equation.  As before,
the constraints are enforced by requirements that can be expressed as
the functional differential equations
\begin{equation} 
\frac{\delta S}{\delta N} = \frac{\delta S}{\delta N_i} = 0 ,
\end{equation} 
or equivalently (cf.~section \ref{IV})
\begin{equation} 
\label{hscnsD}
 \HH\left( q_A, \frac{\delta S}{\delta q_I}, p^\mu \left( q_A, 
\dot{q}_I\left( q_A, \frac{\delta S}{\delta q_I}, 
\dot{q}_\mu, N, N_i\right) , \dot{q}_\mu , N, N_i\right) , N, 
N_i\right) = 0 .
\end{equation} 
\begin{equation} 
\label{hiscnsD}
\HH^i\left( q_A, \frac{\delta S}{\delta q_I}, p^\mu \left( q_A, 
\dot{q}_I\left( q_A, \frac{\delta S}{\delta q_I}, 
\dot{q}_\mu, N, N_i\right) , \dot{q}_\mu , N, N_i\right) , N, 
N_i\right) = 0 
\end{equation}  
On the geometrodynamic superspace we have, in general, 
\begin{equation} 
\label{STD}
\frac{\delta S}{\delta t} \not = 0 
\end{equation} 
even on the constraints shell. The reason for this is that, unlike the 
variational derivative of $S$ with respect to $t$ of equation (\ref{STG}) 
computed at fixed $q_A$, the derivative of equation (\ref{STD}) is 
computed at the fixed true dynamic variables $q_I$ only. 

As before, there are two ways to solve the equations for $S$, similar
to those described above for the configuration superspace of $q_A$.
The analog of the traditional choice would be to solve first the
system of equations (\ref{hjeqD}) and (\ref{hiscnsD}), which will
partially fix the form of $S$.  After that, solve (\ref{hscnsD}),
which is considered the proper Hamilton--Jacobi equation.  However,
such a choice mixes the true dynamical variables with embedding
parameters at both stages and looses all the advantages of the similar
procedure on the superspace of all $q_A$.  This goes against the
structure of the theory and implies the loss of any similarity to the
Hamilton--Jacobi theory in mechanics.  The second approach is to solve
(\ref{hjeqD}) first, considering it the Hamilton--Jacobi equation, and
then using (\ref{hscnsD}) and (\ref{hiscnsD}) to adjust the functional
parameters of the solution.  The first stage of the procedure
provides, in principle, solutions of the Hamilton--Jacobi equation
both on and off shell, and thus includes information that disappears
at the second stage when the solution is forced onto the constraint
shell. In practice, this procedure is more involved than it appears,
but in the end is equivalent to the procedure on the superspace of
$q_A$ described above.

To summarize, the Hamilton--Jacobi theories on the superspace of
3--metrics (parametrized by $g_{ik}$ or $q_A$) and on the
geometrodynamic superspace are found to be equivalent.  Any possible
difference is erased by forcing solutions onto the constraint shell.
Identification of the Hamilton--Jacobi equation is, to an extent,
arbitrary.  The total content of the Hamilton--Jacobi theory is
expressed by three equations, be it on the superspace of 3--geometries
(equations (\ref{hjeqgq}), (\ref{hscnsq}), and (\ref{hiscnsq})) or on
the geometrodynamic superspace (equations (\ref{hjeqD}),
(\ref{hscnsD}), and (\ref{hiscnsD})).  Identification of one of the
equations as the Hamilton--Jacobi equation is related to the choice of
solution strategy and does not influence the final solution.  It is
mostly a matter of convenience and, as such, is problem dependent.
The situation changes when the theory is quantized because it is the
Hamilton--Jacobi equation that is converted into the wave equation,
while the other two equations merely supply additional information.

\section{Canonical Quantization} 
\label{VI} 

Canonical quantization of any field theory is based on the
Hamilton--Jacobi representation of the theory and consists of steps
that are determined by this representation.  Instead of the classical
system determined by the functional $S$, a quantum system determined
by the state functional $\Psi$ is introduced.  The arguments of $\Psi$
are assumed to be the same as the arguments of $S$, and are determined
by the choice of the configuration superspace and by the assignment of
the equation to be treated as the Hamilton--Jacobi equation.

The classical Hamilton--Jacobi equation is then transformed into the
wave equation of the quantum theory.  The functional derivatives of
$S$ in the Hamilton--Jacobi equation are replaced by operators acting on
$\Psi$ as follows
\begin{eqnarray} 
-\frac{\delta S}{\delta t} & \Longrightarrow &\quad
i\hbar \frac{\delta}{\delta t}  \\
\frac{\delta S}{\delta g_{ik}} & \Longrightarrow &\quad
\widehat{\pi}^{ik} =\ \frac{\hbar}{i}\,  
\frac{\delta}{\delta g_{ik}} \\
\frac{\delta S}{\delta q_A} &\Longrightarrow &\quad
\widehat{p}^A  =\  \frac{\hbar}{i}\,  
\frac{\delta}{\delta q_A} \\
\frac{\delta S}{\delta q_I} &\Longrightarrow &\quad
\widehat{p}^I =\ \frac{\hbar}{i}\,  
\frac{\delta}{\delta q_I}.
\end{eqnarray} 
The rest of the expressions participating in the Hamilton--Jacobi
equations are interpreted as c--numbers acting either on the density
$\psi$ of $\Psi$, defined by
\begin{equation} 
\Psi = \int \psi\, d^3x ,
\end{equation} 
or the densities produced by functional differentiation of $\Psi$.
For instance, if $A$ is an expression that does not contain momenta,
the action of the operator $\widehat A$ associated to it will be given
by
\begin{eqnarray} 
\widehat A\, \Psi &= & A\, \psi \\ 
\widehat A\, \frac{\delta\Psi}{\delta q_A} &=& 
A\, \frac{\delta\Psi}{\delta q_A} \\ 
\widehat A\, \frac{\delta^2\Psi}{\delta q_A \delta q_B} &=& 
A\, \frac{\delta^2\Psi}{\delta q_A \delta q_B} 
\end{eqnarray} 
and so on. 

On the configuration superspace of 3--metrics, with equation
(\ref{hscns}) identified as the Hamilton--Jacobi equation, this
procedure produces the Hamilton operator (we ignore here technical
problems such as factor ordering)
\begin{equation} 
\widehat\HH = \widehat\HH\left( g_{ij}, \widehat{\pi}^{ij}\right) 
\end{equation} 
and the wave equation  
\begin{equation} 
\widehat\HH\left( g_{ij}, \widehat{\pi}^{ij}\right)\, \Psi = 0,
\end{equation} 
known as the Wheeler--DeWitt equation.

Similarly, the change of variables that introduces the superspace of
$q^A$, and the assignment of (\ref{hscnsq}) as the Hamilton--Jacobi
equation, produces the Hamilton operator
\begin{equation}
\widehat\HH = \widehat\HH\left( q_A, \widehat{p}^A\right)  
\end{equation} 
and the wave equation 
\begin{equation}
\widehat\HH\left( q_A, \widehat{p}^A\right)\, \Psi = 0   .
\end{equation} 
A more detailed form of this (Wheeler--DeWitt) equation, based on
(\ref{HJGq}), is
\begin{equation}  
-\hbar^2\, g^{-\frac{1}{2}}\, Q_{AB}\, 
\frac{\delta^2 \Psi}{\delta q_A\, \delta q_B} 
- g^{\frac{1}{2}}\, \widehat R\, \Psi = 0   ,
\end{equation} 
which is a second order functional differential equation.  The second
functional derivatives do not participate in the classical
Hamilton--Jacobi theory; simple descriptions of these operations can
be found in the literature\cite{Gelf} (more sophisticated treatments
are also easy to find).  This equation does not contain the time
derivative of $\Psi$.  For it to describe the time evolution of
$\Psi$, time must be inserted into the equation by assigning a
suitable function of the 3--metric.  The resulting equation cannot be
interpreted as a Schr\"odinger equation. Its structure is more similar
to that of the Klein--Gordon equation, and this is the cause of some
of the problems of time in the quantum picture associated with this
approach.

The wave equation constitutes only a part of the theory.  In addition,
commutation relations are imposed on the coordinates of the
configuration superspace ($g_{ij}$ or $q_A$) and their conjugate
momenta ($\widehat{\pi}^{ij}$ or $\widehat{p}^A$). The state
functional is a distribution over the superspace of configurations,
which makes inserting time into the quantum picture, as described
above, more troublesome than it initially appears.

The quantum version of the auxiliary relations (\ref{hiscns}) and
(\ref{hiscnsq}) (supermomentum constraints) are obtained by forming
the operators
\begin{equation} 
\widehat{\HH}^i = \widehat{\HH}^i\left( g_{ij}, \widehat{\pi}^{ij}\right)  
\end{equation}  
or
\begin{equation} 
\widehat{\HH}^i = \widehat{\HH}^i\left( q_A, \widehat{p}^A\right)  
\end{equation} 
and enforcing the constraints by operator equations that can be 
written, for both cases, as 
\begin{equation} 
\widehat{\HH}^i\, \Psi = 0 .
\end{equation} 
In the case of the configuration superspace of $q_A$ this equation is
obtained from (\ref{dinvq}), and can be written as (again, ignoring
factor ordering issues)
\begin{equation}  
\left( -2\, \frac{\hbar}{i}\, Q_{AB}\, \frac{\delta \Psi}{\delta q_A}\, 
{M^B}_{lm}\, G^{lm\, ij}\right)_{|j} = 0    .
\end{equation} 
The tendency to enforce the supermomentum constraints as operator
equations, after the superhamiltonian constraint has been interpreted
as the wave equation and written as an operator equation, is quite
understandable.  After all, the classical versions of these equations
are merely different components of an equation that ensures
energy--momentum conservation.  However, once implemented, these
equations generate the problems of time and prevent the introduction
of any concept of time or of geometrodynamic evolution\cite{Kuc92}.

Comparing this quantum gravity picture with quantum electrodynamics,
we find that the equations in question play a role similar to that of
the Lorentz gauge condition (charge conservation).  In quantum
electrodynamics the Lorentz gauge is imposed as a statement concerning
expectation values rather than as an operator equation\cite{Heit}. A
similar approach in quantum gravity does not seem to be viable in the
picture based on the Wheeler--DeWitt equation.

In contrast, canonical quantization on the geometrodynamic superspace
of $q_I$, with equation (\ref{hjeqD}) identified as Hamilton--Jacobi
equation, appears to produce a quantum gravitational picture that
avoids the problems of time, and which is similar in spirit to quantum
electrodynamics.  The geometrodynamic quantum Hamiltonian operator is
based on the expression for $\HH_{DYN}$ on the right hand side of
equation (\ref{hjeqD}),
\begin{equation}
\widehat{\HH}_{DYN} = 
 \widehat{\HH}_{DYN} \left( q_A, 
\widehat{p}^I, p^\mu \left( q_A, 
\dot{q}_I\left( q_A, \widehat{p}^I, \dot{q}_\mu, 
N, N_i\right) , \dot{q}_\mu , N, N_i\right) , N, N_i\right),
\end{equation} 
where $\widehat{p}^I$ is given by 
\begin{equation} 
\widehat{p}^I = \frac{\hbar}{i}\, \frac{\delta}{\delta q_I} .
\end{equation} 
The wave equation of the geometrodynamic quantum theory (Schr\"odinger 
equation),
\begin{equation} 
\label{SCHR}
-i \hbar \frac{\delta \Psi}{\delta t} = \widehat{\HH}_{DYN}\, \Psi   ,
\end{equation} 
is obtained from (\ref{hjeqD}) in the standard way.   

The commutation relations are imposed only on the true dynamical
variables $q_I$, and their conjugate momenta $\widehat{p}^I$.
Embedding variables and their velocities are c--numbers, and as such
generate only trivial commutation relations.  Time, suitable for
describing quantum geometrodynamic evolution, can easily be introduced
through these embedding variables.

The auxiliary conditions (\ref{hscnsD}) and (\ref{hiscnsD}) of the
classical theory are replaced by quantum conditions, based on
operators derived from the left hand side of these equations.  This is
achieved using a procedure similar to the one used to form
$\widehat{\HH}_{DYN}$:
\begin{equation} 
\label{HMHAT}
\widehat{\HH} = \widehat{\HH}\left( q_A, \widehat{p}^I, p^\mu \left( q_A, 
\dot{q}_I\left( q_A, \widehat{p}^I, 
\dot{q}_\mu, N, N_i\right) , \dot{q}_\mu , N, N_i\right), N, 
N_i\right) 
\end{equation}        
\begin{equation} 
\label{SMHAT} 
\widehat{\HH}^i = \widehat{\HH}^i\left( q_A, \widehat{p}^I, p^\mu \left( q_A, 
\dot{q}_I\left( q_A, \widehat{p}^I, 
\dot{q}_\mu, N, N_i\right) , \dot{q}_\mu , N, N_i\right), N, 
N_i\right)  .
\end{equation}
  
It has been mentioned above that imposing the constraints in operator
form leads to numerous difficulties, including time problems. A weaker
way to impose the constraints, similar to that of quantum
electrodynamics, is to impose the constraints on expectation values.

The first step of this procedure is to solve the Schr\"odinger
equation (\ref{SCHR}), with appropriate initial and boundary
conditions, assuming that embedding variables are represented by
c--numbers which are unknown but assumed to exist. The resulting
solution $\Psi_s$ is a functional that can be represented as
\begin{equation} 
\Psi_s(t, q_I) = \int \psi_s(t, q_I; x^i)\, d^3x .
\end{equation} 

The action of any observable $\widehat A$ on this produces a function
on slices.  The expectation of this observable over the solution can
be written symbolically as
\begin{equation} 
A_s = \left\langle \Psi_s |\widehat A| \Psi_s \right\rangle 
\end{equation} 
and computed following the prescription 
\begin{equation} 
A_s = \int \psi^*_s\, \widehat A\, \Psi_s\, \mathcal{D}q_I\ \mathcal{D}q_J ,
\end{equation} 
which includes functional integration over the geometrodynamic
configuration superspace but not over slices.

Quantum constraints on the level of expectations can then be formed as
\begin{eqnarray} 
\label{expH} 
\left\langle \Psi_s | \widehat\HH | \Psi_s \right\rangle &=& 0 \\
\label{expHI}
\left\langle \Psi_s | \widehat{\HH}^i | \Psi_s \right\rangle &=& 0 ,
\end{eqnarray}   
with $\widehat\HH$ and $\widehat{\HH}^i$ given by (\ref{HMHAT}) and 
 (\ref{SMHAT}).

\section{Discussion.} 
\label{VII} 
 
In this paper we have extended the
geometrodynamic quantization approach from quantum cosmological
models\cite{KheMil94a} to the generic gravitational field.  We find no
inconsistencies in this broader setting.  Quantum geometrodynamic
evolution is determined by the Schr\"odinger equation (\ref{SCHR}),
together with the constraints (\ref{expH}) and (\ref{expHI}) imposed
on expectations.  Lapse and shift should be either given explicitly or
determined by four additional conditions, which determine the
interpretation of the time parameter $t$.  If the classical version of
the conditions includes the true dynamical variables and their
conjugate momenta, their quantum version is imposed in the form of
expectations, just like the constraints.

Solving any particular problem can be thought of as a three--step
procedure.  First, the Schr\"odinger equation (\ref{SCHR}) is solved,
assuming that the embedding variables and their time derivatives are
unique, although unknown, functions of time and the spatial
coordinates.  The same is assumed of lapse and shift, unless they are
given explicitly.  Solving the the Schr\"odinger equation implies that
appropriate boundary and initial conditions for the state functional
on the geometrodynamic configuration superspace are given.  The
resulting solution is a functional that depends on the embedding
variables and their time derivatives, as well as shift and lapse.  The
expectations of the constraints (\ref{expH}) and (\ref{expHI}) over
the solution of the Schr\"odinger equation are then computed.  This
procedure produces four differential equations for the four embedding
variables if the lapse and shift are given explicitly.  Alternatively,
one can simply couple the four constraint conditions with four
functional conditions for the lapse and shift.  These procedures
determine the meaning of time. The last step is to solve these
equations and substitute the solutions for the embedding variables,
their time derivatives, and lapse and shift into the expression for
the state functional.

This whole quantization procedure for the general gravitational field
parallels the quantum cosmological examples considered
elsewhere\cite{KheMil94a}.  Considerable complications are caused by
the algebraic complexity of the expressions for the geometrodynamic
Hamiltonian and the constraints, as well as by the functional nature
of equations.  These complications do not, however, stop the solution
procedure in principle, although they do introduce a rather complex
coupled system.  This complexity places demands on our ability to gain
a proper understanding of the problem, especially in setting
appropriate initial and boundary conditions on the configuration
superspace.

In principle, however, the geometrodynamic quantization formalism in
the general setting retains all of the essential features previously
illustrated in the context of homogeneous cosmologies.

\end{document}